\documentclass[fleqn,12pt]{article}
\usepackage{epsfig}
\begin{document}
\textheight 22cm
\textwidth 15cm
\noindent
{\Large \bf Comparison of multi-scale analysis models applied to zonal flow generation in ion-temperature-gradient mode turbulence}
\newline
\newline
J. Anderson\footnote{anderson.johan@gmail.com}, Y. Kishimoto
\newline
Department of Fundamental Energy Science
\newline
Graduate School of Energy Science, Kyoto University, Gokasho, Uji, Kyoto 611-0011
\newline
\newline
\begin{abstract}
\noindent
During the past years the understanding of the multi-scale interaction problems have increased significantly. However, at present there exists a range of different analytical models for investigating multi-scale interactions and hardly any specific comparisons have been performed among these models. In this work, two different models for the generation of zonal flows from ion-temperature-gradient (ITG) background turbulence are discussed and compared. The methods used is the coherent mode coupling model and the wave kinetic equation model (WKE). It is shown that the two models give qualitatively the same results even though the assumption on the spectral difference is used in the (WKE) approach.
\end{abstract}
\newpage
\renewcommand{\thesection}{\Roman{section}}
\section{Introduction}
\indent
The understanding of multi-scale problems is believed to be a key to gain deeper knowledge in basic plasma physics. The multi-scale analysis have had great success during the last couple of years and been able to give great insight in the complex plasma physical processes in e.g. the zonal flow - drift wave system~\cite{a11}. There are a large number of theoretical-analytical studies devoted to multi-scale problems using different background physics Refs~\cite{a12}-~\cite{a26} and there have also been intensive numerical investigations using gyrokinetic~\cite{a27}-~\cite{a291} and advanced fluid models~\cite{a30}-~\cite{a35} of the the zonal flow - drift wave system.

The problem of anomalously high ion heat transport believed to be generated by the ion-temperature-gradient (ITG) mode turbulence and controlled by the secondary generated zonal flows is one of the typical multi-scale problems~\cite{a12}-~\cite{a14}. The zonal flows are azimuthally symmetric shear flows with temporal variation with time scales longer than the drift waves and having disparate spatial scale lengths. This is often utilized to reduce the fully coupled system to an analytically tractable problem. It is most plausible that robust quasi-stationary zonal flows may suppress the ion heat transport and give access to improved confinement regimes in tokamak plasmas. Thus, the understanding of the dynamical system of zonal flow - drift waves is of great importance for controlling the anomalous transport and the formation of internal transport barriers (ITB). 

The zonal flow - drift wave system has been studied previously using multi-scale analysis for different physics such as the electromagnetic ITG mode~\cite{a19}, the trapped-electron-mode (TEM)~\cite{a25}, the electron-temperature-gradient (ETG) mode~\cite{a191},~\cite{a231} etc. In the case of TEM background turbulence the generation of zonal flows seem to be much weaker than for the electrostatic ITG mode~\cite{a25} and also for the ETG mode the zonal flow instability is quite weak. It has been speculated that the importance of zonal flows is of less interest in ETG turbulence. 

The problem of multi-scale interaction between a tearing mode and drift waves has just recently been discussed~\cite{a26}.

There are several analytical models for treating multi-scale interactions and among the widely used models are the coherent mode coupling method~\cite{a16},~\cite{a19}-~\cite{a192}, the wave kinetic equation approach~\cite{a15},~\cite{a17}-~\cite{a18},~\cite{a21},~\cite{a23}-~\cite{a25},~\cite{a26} and the reductive perturbation expansion method~\cite{a20},~\cite{a251}-~\cite{a252}. There is indeed little comparative knowledge of the differences in the current methods used.

The main purpose of the paper is to compare and point out differences in two of the main approaches to treat multi-scale problems. The comparison is done by deriving a model for a zonal flow generated from background ITG turbulence using the method of coherent mode coupling and compare this to an existing model derived using the Wave Kinetic Equation (WKE)~\cite{a18},~\cite{a24} using the same background physics and assumptions. Direct Numerical Simulations has been performed, showing that the scale separation assumed in the WKE model breaks down during the simulations~\cite{a351} (and References therein). The method proposed here and also briefly discussed in Ref.~\cite{a192} are complementing this previous line of research to find deeper understanding for the approximations made in multi-scale modeling. 

In this paper, an algebraic dispersion relation for zonal flow growth rate and real frequency is derived using the coherent mode coupling method in the presence of toroidal ITG turbulence. The model for the background turbulence is based on the ion-continuity and the ion-temperature-equation. The evolution of the zonal flow is described by a Hasegawa - Mima like equation. The models employed are electrostatic and effects of trapping are neglected.   

It is found that the two different models for estimating the effects of multi-scale interaction, in this particular case the interaction of ITG driven drift waves and a zonal flow, are in good quantitative and qualitative agreement in the region of maximum linear drift wave growth rate (close to $k_x \rho = k_y \rho \approx 0.3$). In all cases good qualitative agreement was found.

The paper is organized as follows. In Section II the analytical model for the zonal flows generated from toroidal ITG modes is reviewed. Section III is dedicated to the results and a discussion thereof. Finally there is a summary in section IV.

\section{Analytical models for zonal flow generation}
The model for the toroidal ITG driven modes consists of the ion continuity and ion temperature equations~\cite{a36}. In the present work the effects of magnetic shear, trapped particles and finite beta is neglected. It has been found earlier that the effect of parallel ion motion on the ITG mode is rather weak~\cite{a36}, however, it is recently found that magnetic shear may modify the zonal flow generation significantly~\cite{a20}. The effects of magnetic shear is out of the scope of the present paper. In this section the model using the coherent mode coupling is presented. The method how to construct the interaction of a zonal flow and the ITG mode turbulence and the derivation of the dispersion relation for zonal flows follows earlier papers~\cite{a16},~\cite{a19}-~\cite{a191} (and References therein). 
 
\begin{eqnarray}
\frac{\partial \tilde{n}}{\partial t} - \left(\frac{\partial}{\partial t} - \alpha_i \frac{\partial}{\partial y}\right)\nabla^2_{\perp} \tilde{\phi} + \frac{\partial \tilde{\phi}}{\partial y} - \epsilon_n g \frac{\partial}{\partial y} \left(\tilde{\phi} + \tau \left(\tilde{n} + \tilde{T}_i \right) \right) = \nonumber \\
- \left[\phi,n \right] + \left[\phi, \nabla^2_{\perp} \phi \right] + \tau \left[\phi, \nabla^2_{\perp} \left( n + T_i\right) \right] \\
\frac{\partial \tilde{T}_i}{\partial t} - \frac{5}{3} \tau \epsilon_n g \frac{\partial \tilde{T}_i}{\partial y} + \left( \eta_i - \frac{2}{3}\right)\frac{\partial \tilde{\phi}}{\partial y} - \frac{2}{3} \frac{\partial \tilde{n}}{\partial t} = \nonumber \\
- \left[\phi,T_i \right] + \frac{2}{3} \left[\phi,n \right].
\end{eqnarray}
The system is closed by using quasi-neutrality with Boltzmann distributed electrons. The slowly varying zonal flow mode is generated from a Hasegawa-Mima type of equation
\begin{eqnarray}
-\frac{\partial}{\partial t} \nabla_x^2 \phi_{ZF} -\mu \nabla_x^4 \phi_{ZF} = \left[\phi, \nabla_{\perp}^{2} \phi\right].
\end{eqnarray}
Here $\left[ A ,B \right] = \partial A/\partial x \partial B/\partial y - \partial A/\partial y \partial B/\partial x$ is the Poisson bracket. With the additional definitions $\tilde{n} = \delta n / n_0$, $\tilde{\phi} = e \delta \phi /T_e$, $\tilde{T}_i = \delta T_i / T_{i0}$ as the normalized ion particle density, the electrostatic potential and the ion temperature, respectively. In the forthcoming equations $\tau = T_i/T_e$, $\vec{v}_{\star} = \rho_s c_s \vec{y}/L_n $, $\rho_s = c_s/\Omega_{ci}$ where $c_s=\sqrt{T_e/m_i}$, $\Omega_{ci} = eB/m_i c$. We also define $L_f = - \left( d ln f / dr\right)^{-1}$, $\eta_i = L_n / L_{T_i}$, $\epsilon_n = 2 L_n / R$ where $R$ is the major radius and $\alpha_i = \tau \left( 1 + \eta_i\right)$. The perturbed variables are normalized with the additional definitions $\tilde{n} = (L_n/\rho_s) \delta n / n_0$, $\tilde{\phi} = (L_n/\rho_s e) \delta \phi /T_e$, $\tilde{T}_i = (L_n/\rho_s) \delta T_i / T_{i0}$ as the normalized ion particle density, the electrostatic potential and the ion temperature, respectively. The perpendicular length scale and time are normalized to $\rho_s$ and $L_n/c_s$, respectively. The geometrical quantities are calculated in the strong ballooning limit ($\theta = 0 $, $g\left(\theta = 0, \kappa \right) = 1/\kappa$ (Ref.~\cite{a38}) where $g\left( \theta \right)$ is defined by $\omega_D \left( \theta \right) = \omega_{\star} \epsilon_n g\left(\theta \right)$).
The linear solutions to Eqs 1 and 2 is,
\begin{eqnarray}
\omega_r & = & \frac{k_y}{2\left( 1 + k_{\perp}^2\right)} \left( 1 - \left(1 + \frac{10\tau}{3} \right) \epsilon_n g - k_{\perp}^2 \left( \alpha_i + \frac{5}{3} \tau \epsilon_n g \right)\right)  \\
\gamma & = & \frac{k_y}{1 + k_{\perp}^2} \sqrt{\tau \epsilon_n g\left( \eta_i - \eta_{i th}\right)}.
\end{eqnarray}
where $\omega = \omega_r + i \gamma$ and 
\begin{eqnarray}
\eta_{i th} \approx \frac{2}{3} - \frac{1}{2 \tau} + \frac{1}{4 \tau \epsilon_n g} + \epsilon_n g\left( \frac{1}{4 \tau} + \frac{10}{9 \tau}\right).
\end{eqnarray}
FLR effects in the $\eta_{i th}$ are neglected.

Following the same procedure as in Refs~\cite{a16},~\cite{a19}-~\cite{a191} by choosing a monochromatic pump wave that generates a secondary convective cell,
\begin{eqnarray}
\tilde{\xi}_{DW}(x,y,t) & = & \xi_{DW} e^{i(k_x x + k_y y - \omega t)} + c.c. \\
\tilde{\xi}_{ZF}(x,y,t) & = & \xi_{ZF} e^{i(q_x x - \Omega t)} + c.c. \\
\tilde{\xi}_\pm(x,y,t) & = & \xi_{\pm}e^{i((q_x \pm k_x)x \pm k_y y - \omega_{\pm }t)} + c.c.
\end{eqnarray}
Here $\tilde{\xi}_{DW}$ is the drift pump wave ($\tilde{n}_i, \tilde{\phi}, \tilde{T}_i$), $\tilde{\xi}_{ZF}$ is the secondary generated wave or zonal flow ($\tilde{\phi}, \tilde{T}_i$ remembering that the zonal flow does not have a density component), $\tilde{\xi}_\pm$ are the possible sidebands ($\tilde{n}_i, \tilde{\phi}, \tilde{T}_i$), $(k_x,k_y)$ are the radial and poloidal wave numbers for the background turbulence, $\omega$ is the complex frequency of the drift wave, $q_x$ is the radial wave number, $\Omega$ is the complex frequency of the zonal flow. Where the frequencies couples as follows $\omega_{\pm} = \Omega \pm \omega$. In Figure 1 the spectral relationship between the waves is shown.

\begin{figure}
  \includegraphics[height=.3\textheight]{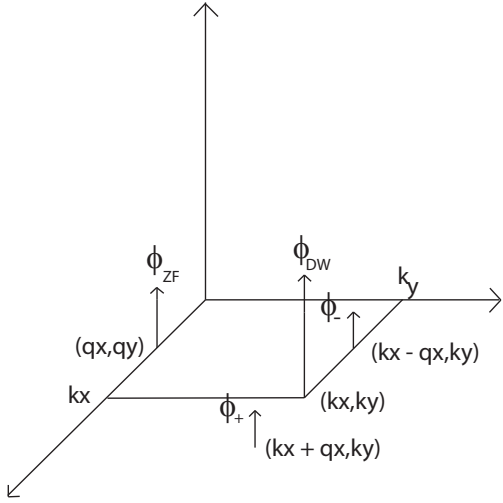}
  \caption{The position of the drift wave pump ($\phi_{DW}$), the side bands ($\phi_{+},\phi_{-}$) and the zonal flow seed ($\phi_{ZF}$) in wave number space. The zonal flow seed is further assumed to have $q_y = 0$.}
\end{figure}

This gives a problem with four coupled waves where the zonal flow seed may be unstable and grow exponentially. Following the method developed in previous work Ref. ~\cite{a16} the sidebands may couple to the pump wave to give the zonal flow component. The sideband potentials are then found using Eq. 1 and identifying the coupling between the drift wave component and zonal flow component as the sideband component. This gives the equation system,
\begin{eqnarray} 
(i\Omega q_x^2 - \mu q_x^4) \phi_{ZF} & = & k_y q_x^2 [a_{+} \phi_{DW}^{*} \phi_{+} + a_{-}\phi_{DW} \phi_{-}] \\
\phi_{+} & = &  i\beta q_x k_y k_{\perp -}\frac{\phi_{DW} \phi_{ZF}}{\Omega + \alpha} \\
\phi_{-} & = & - i\beta q_x k_y k_{\perp -}\frac{\phi_{DW}^{*} \phi_{ZF}}{\Omega - \alpha}
\end{eqnarray}
In deriving these equations the FLR effects are considered to be small. Here 
\begin{eqnarray} 
a_{\pm} & = & 2 k_x \pm q_x \\
k_{\perp -} & = & k_x^2 + k_y^2 - q_x^2 \\
\beta & = & 1 + \tau + \tau \delta \\
\delta & = & \frac{(\eta_i - \frac{2}{3})k_y + \frac{2}{3}\omega}{\omega + \frac{5}{3}\tau \epsilon_n g k_y} \\
 \alpha & = & \omega - k_y + \epsilon_n g k_y \beta
\end{eqnarray}
It is now possible to derive a 3rd order dispersion relation for the zonal flow component.
\begin{eqnarray}
(\Omega + i \mu q_x^2)(\Omega^2 - \alpha^2) =  2 \beta k_y^2 q_x k_{\perp -} [ q_x \Omega - 2 k_x \alpha] |\phi_{DW}|^2
\end{eqnarray}
This third order dispersion relation is solved numerically and compared to the basic result for zonal flows below Ref.~\cite{a24}.

The dispersion relation for the zonal flow derived using the wave kinetic model is thoroughly described in Ref.~\cite{a24} and hence only the final result is given here. The zonal flow dispersion relation is found using the vorticity equation and combining this with the wave kinetic equation as an additional relation between the background turbulence and the zonal flow. In describing the large scale plasma flow dynamics it is assumed that there is a sufficient spectral gap between the small scale fluctuations and the large scale flow i.e. for the poloidal wavenumber it is assumed that $q_y<<k_y$ and the wave kinetic equation is solved in the radial direction. It is also assumed that there is enough difference in the temporal variation of micro turbulence and zonal flow ($\Omega_{ZF}<<\omega_{DW}$). The electrostatic potential is represented as a sum of fluctuating and mean quantities. In the case of zonal flows ($\Omega$) generated from ITG mode turbulence the dispersion relation is
\begin{eqnarray}
\left(\Omega + i \mu q_x^2 \right)\left(\Omega - q_x v_{gx}\right)^2 = - q_x^2 \left(1 + \tau  + \tau \delta \right) |\phi_{DW}|^2 \Omega 
\end{eqnarray}
Here $v_{gx} = \frac{\partial \omega_r}{\partial k_x}$ is the group velocity for the drift wave and the other parameters are the same as above. An alternative statistical approach, resulting in a modified wave kinetic equation, is presented in Ref.~\cite{a21} and~\cite{a211} which also contains an extensive discussion of and comparison with the approach used here. 

\section{Results and discussion}
The growth rates and real frequencies from the coherent mode coupling model (Eq. 18) and the WKE model (Eq 19) are compared for a variety of plasma parameters. The dispersion relations are solved numerically and all growth rates and real frequencies are compared to the corresponding linear ITG growth rate (Eq. 5) except where explicitly stated. In this study we have assumed that the mode coupling saturation level ($|\phi_{DW}| = \frac{1}{k_x L_n} \frac{\gamma}{\omega_*}$)~\cite{a40} is reached for the background turbulence.

First in Figure 2 and 3 the zonal flow growth rate (positive sign) and real frequency (negative sign) as a function of $\epsilon_n$ with $\eta_i$ as a parameter is shown for the coherent mode coupling model and the WKE model, respectively. The parameters are $\tau = 1$, $k_x = k_y = q_x = 0.3$, $\mu = 0$, $\eta_i = 3.0$ (square), $\eta_i = 4.0$ (asterisk) $\eta_i = 5.0$ (diamond). In the figures it is found that the zonal flow growth rate is not very sensitive to a change in $\eta_i$ on the other hand in both figures large zonal flow growth rates are found except for small $\epsilon_n$ (peaked density profiles). In the models large levels of zonal flow generation ($\gamma_{ZF}>\gamma_{ITG}$) is displayed for a rather large parameter regime. Moreover, in the variation of $\epsilon_n$ quite good qualitative and quantitative agreement for the growth rates and real frequencies in both figures are displayed. The quantitative agreement of in both cases are dependent on $q_x$ c.f Figure 6, whereas, the qualitative agreement is conserved. 

\begin{figure}
  \includegraphics[height=.3\textheight]{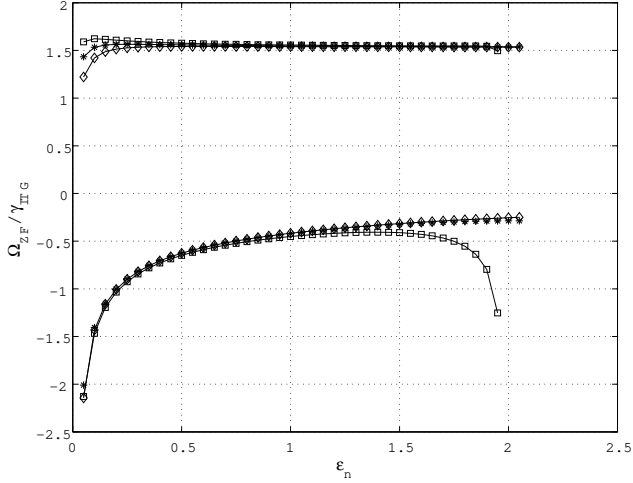}
  \caption{The zonal flow growth rate and frequency (normalized to the ITG mode growth) as a function of $\epsilon_n$ with $\eta_i$ as a parameter for the coherent mode coupling model model. The parameters are $\tau = 1$, $k_x = k_y = q_x = 0.3$, $\mu = 0$, $\eta_i = 3.0$ (square), $\eta_i = 4.0$ (asterisk) $\eta_i = 5.0$ (diamond).}
\end{figure}

\begin{figure}
  \includegraphics[height=.3\textheight]{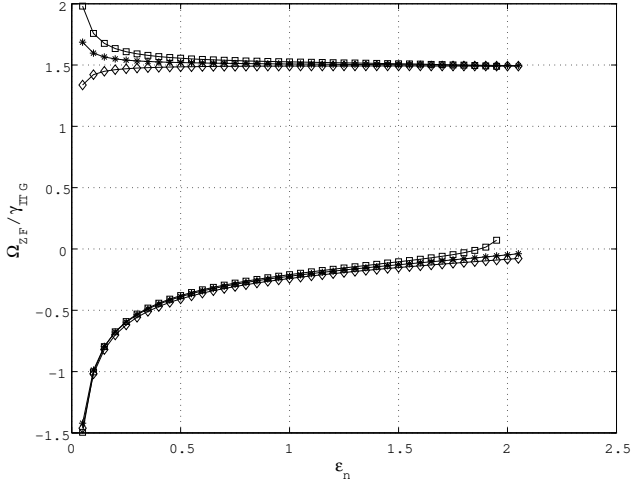}
  \caption{The zonal flow growth rate and frequency (normalized to the ITG mode growth) as a function of $\epsilon_n$ with $\eta_i$ as a parameter for the WKE model. The parameters are $\tau = 1$, $k_x = k_y = q_x = 0.3$, $\mu = 0$, $\eta_i = 3.0$ (square), $\eta_i = 4.0$ (asterisk) $\eta_i = 5.0$ (diamond).}
\end{figure}
 
Second, the effect of a variation of the background turbulence spectrum ($k_x,k_y$) on the zonal flow growth rate (normalized to $c_s/L_n$) is shown in Figure 4 and 5. The zonal flow growth rate as a function of $k_y$ with $k_x$ as a parameter is displayed for the coherent mode coupling model and the WKE model, respectively. The parameters are $\eta_i = 4$, $\tau = 1$, $q_x = 0.3$, $\mu = 0$ and $k_x = 0.2$ (asterisk), $k_x = 0.3$ (square), $k_x = 0.4$ (plus). The comparison is done for relatively small values of ($k_x,k_y$) since the perpendicular wave vector is considered small $k_{\perp}^2 << 1 $ in the derivations of the dispersion relations for the zonal flow growth rate and real frequency. The zonal flow growth rates are increasing with increasing $k_y$ (strong zonal flow generation is found if streamer like pump is assumed) and have a rather modest dependency on $k_x$. The growth rates are decreasing for increasing $k_x$. The zonal flow growth rates found for the two different models in Figure 3 and 4 are in qualitative agreement and have good quantitative agreement in the ($k_x,k_y$) region where the linear ITG growth rate is at maximum (close to $k_x = k_y = 0.3$). For small $k_x$ there is a deviation in the $k_y$ scaling of the growth rate in the coherent mode coupling model, whereas, for larger $k_x$ the behavior is rather similar. 

\begin{figure}
  \includegraphics[height=.3\textheight]{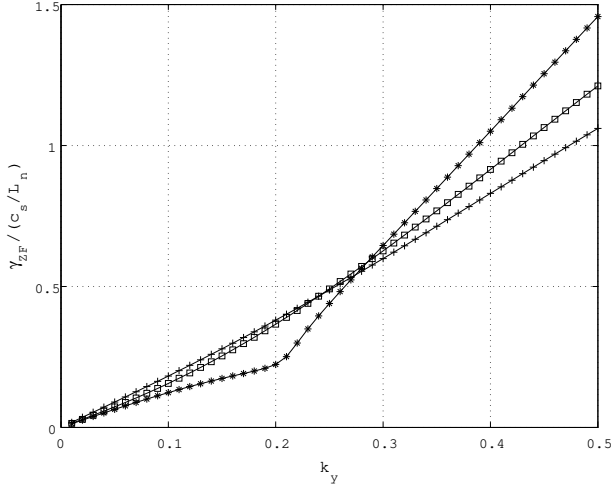}
  \caption{The zonal flow growth rate in the coherent mode coupling model (normalized to $c_s/L_n$) as a function of $k_y$ with $k_x$ as a parameter for $\eta_i = 4$, $\tau = 1$, $q_x = 0.3$, $\mu = 0$ and $k_x = 0.2$ (asterisk), $k_x = 0.3$ (square), $k_x = 0.4$ (plus).}
\end{figure}

\begin{figure}
  \includegraphics[height=.3\textheight]{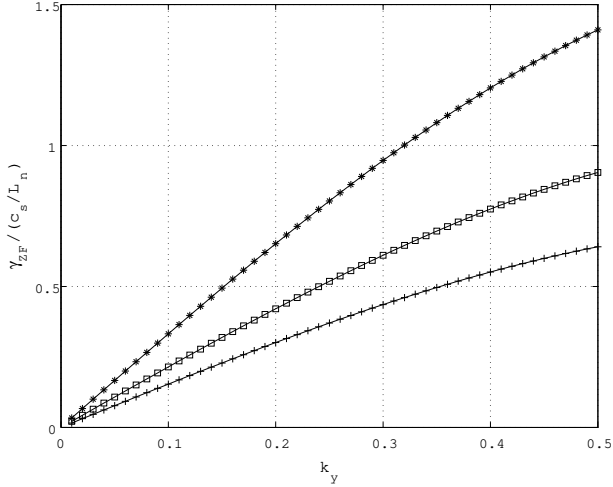}
  \caption{The zonal flow growth rate in the WKE model (normalized to $c_s/L_n$) as a function of $k_y$ with $k_x$ as a parameter for $\eta_i = 4$, $\tau = 1$, $q_x = 0.3$, $\mu = 0$ and $k_x = 0.2$ (asterisk), $k_x = 0.3$ (square), $k_x = 0.4$ (plus).}
\end{figure}

Third, the dependency of the radial wave number of the zonal flow is compared for the WKE (square) and the coherent mode coupling model (asterisk) in Figure 6. The parameters are $\eta_i = 4$, $\epsilon_n = 1.0$, $\tau = 1$, $k_x = k_y = 0.3$, $\mu = 0$. In the case of the WKE model the scalings of the growth rate (positive sign) and real frequency (negative sign) are linear. The coherent mode coupling model has a slightly more complicated behavior where a maximum growth rate is found close to $q_x = 0.2$. The two models exhibits completely different scaling in $q_x$, however, the scalings with other parameters are conserved and it is only a change in the levels. In deriving the WKE dispersion relation, a state close to marginal stability was assumed and the feature of a non vanishing zonal flow growth rate for small $q_x$ may be recovered if this assumption is discarded. However, it should be noted that for $q_x = 0.0$ the Reynolds stress is identically zero and the driving term disappears thus the zonal flow growth rate is zero at this particular value. The zonal flow growth rate reach a maximum if the damping term is included, in the WKE model.

\begin{figure}
  \includegraphics[height=.3\textheight]{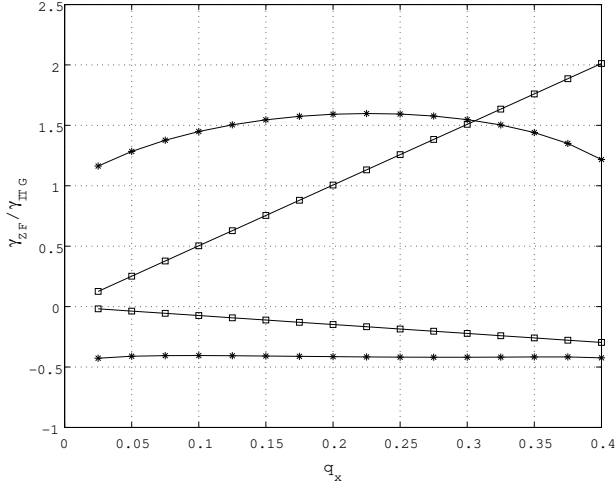}
  \caption{The zonal flow growth rate (normalized to the ITG mode growth) as a function of $q_x$ for the WKE model (squares) and the coherent mode coupling model (asterisks) parameter for $\eta_i = 4$, $\epsilon_n = 1.0$, $\tau = 1$, $k_x = k_y = 0.3$, $\mu = 0$.}
\end{figure}
 
\section{Summary}
The main purpose of the paper is to make a comparison of two of the current methods for studying multi-scale problems. In this work the two methods are employed to study the interaction of a zonal flow with the drift wave system. The first method is based on the wave kinetic equation (model derived in Ref.~\cite{a24}) and the other method is that of coherent mode coupling (derived in this paper). An algebraic dispersion relation for zonal flow growth rate and real frequency is derived using the coherent mode coupling method in the presence of toroidal ITG turbulence. The model for the background turbulence is based on the ion-continuity and the ion-temperature-equation. The evolution of the zonal flow is described by a Hasegawa - Mima like equation. The models employed are electrostatic and effects of trapping are neglected. 

It is found that the zonal flow growth rate is not very sensitive to a change in $\eta_i$. In the models large levels of zonal flow generation ($\gamma_{ZF}>\gamma_{ITG}$) is displayed for a rather large parameter regime. Moreover, in the variation of $\epsilon_n$ quite good qualitative and quantitative agreement for the growth rates for both models are displayed. 

The zonal flow growth rates are increasing with increasing $k_y$ (strong zonal flow generation is found if streamer like pump is assumed) and have a rather modest dependency on $k_x$. The growth rates are decreasing for increasing $k_x$. The zonal flow growth rates found for the two different models in Figure 3 and 4 are in qualitative agreement and have good quantitative agreement in the ($k_x,k_y$) region where the linear ITG growth rate is at maximum. 

It is found that the two different models for estimating the effects of multi-scale interaction, in this particular case the interaction of ITG driven drift waves and a zonal flow, are in good quantitative and qualitative agreement in the region of maximum linear drift wave growth rate (close to $k_x \rho = k_y \rho \approx 0.3$). In all cases good qualitative agreement was found.

To this end, the present work indicates that the assumption of a large spectral difference between the zonal flow mode and the driving background turbulence is of rather small importance. It seems that the important condition is that there is enough difference in time scales of the slow varying zonal flow and the rapidly varying ITG turbulence.
\section{Acknowledgment}
This research was supported by Japan Society for the Promotion of Science (JSPS).

\end{document}